\documentclass[prb,showpacs,twocolumn,preprintnumbers,amsmath,groupedaddress,amssymb,superscriptaddress,floatfix]{revtex4}
\usepackage[dvips]{graphicx}
\usepackage[latin1]{inputenc}
\begin{document}
\draft

\hyphenation{a-long}

\title{Evidence for a vortex-glass transition in superconducting Ba(Fe$_{0.9}$Co$_{0.1}$)$_{2}$As$_{2}$}

\author{G.~Prando}\email[E-mail: ]{g.prando@ifw-dresden.de}\affiliation{Leibniz-Institut f\"ur Festk\"orper- und Werkstoffforschung (IFW) Dresden, D-01171 Dresden, Germany}
\author{R.~Giraud}\affiliation{Leibniz-Institut f\"ur Festk\"orper- und Werkstoffforschung (IFW) Dresden, D-01171 Dresden, Germany}
\author{S.~Aswartham}\affiliation{Leibniz-Institut f\"ur Festk\"orper- und Werkstoffforschung (IFW) Dresden, D-01171 Dresden, Germany}
\author{O.~Vakaliuk}\affiliation{Leibniz-Institut f\"ur Festk\"orper- und Werkstoffforschung (IFW) Dresden, D-01171 Dresden, Germany}
\author{M.~Abdel-Hafiez}\affiliation{Leibniz-Institut f\"ur Festk\"orper- und Werkstoffforschung (IFW) Dresden, D-01171 Dresden, Germany}
\author{C.~Hess}\affiliation{Leibniz-Institut f\"ur Festk\"orper- und Werkstoffforschung (IFW) Dresden, D-01171 Dresden, Germany}
\author{S.~Wurmehl}\affiliation{Leibniz-Institut f\"ur Festk\"orper- und Werkstoffforschung (IFW) Dresden, D-01171 Dresden, Germany}\affiliation{Institut f\"ur Festk\"orperphysik, Technische Universit\"at Dresden, D-01062 Dresden, Germany}
\author{A.~U.~B.~Wolter}\affiliation{Leibniz-Institut f\"ur Festk\"orper- und Werkstoffforschung (IFW) Dresden, D-01171 Dresden, Germany}
\author{B.~B\"uchner}\affiliation{Leibniz-Institut f\"ur Festk\"orper- und Werkstoffforschung (IFW) Dresden, D-01171 Dresden, Germany}\affiliation{Institut f\"ur Festk\"orperphysik, Technische Universit\"at Dresden, D-01062 Dresden, Germany}

\widetext

\begin{abstract}
Measurements of magneto-resistivity and magnetic susceptibility were performed on single crystals of superconducting Ba(Fe$_{0.9}$Co$_{0.1}$)$_{2}$As$_{2}$ close to the conditions of optimal doping. The high quality of the investigated samples allows us to reveal a dynamic scaling behaviour associated with a vortex-glass phase transition in the limit of weak degree of quenched disorder. Accordingly, the dissipative component of the ac susceptibility is well reproduced within the framework of Havriliak-Negami relaxation, assuming a critical power-law divergence for the characteristic correlation time $\tau$ of the vortex dynamics. Remarkably, the random disorder introduced by the Fe$_{1-x}$Co$_{x}$ chemical substitution is found to act on the vortices as a much weaker quenched disorder than previously reported for cuprate superconductors such as, e.~g., Y$_{1-x}$Pr$_{x}$Ba$_{2}$Cu$_{3}$O$_{7-\delta}$.
\end{abstract}

\pacs{74.25.Uv, 74.25.Wx, 74.70.Xa}

\date{\today}

\maketitle

\narrowtext

\section{Introduction}

The possibility of exploring the mixed Shubnikov phase in type-II superconductors over a wide range of external thermodynamic parameters led to important achievements of the research on high-$T_{\textrm{c}}$ materials. The physics of the vortices (or fluxoids, each one bringing a quantized unit of magnetic field flux $\Phi_{0} = hc/2e = 2.068 \times 10^{-7}$ G cm$^{2}$) is indeed of great interest on the fundamental level, the system being modeled as an ensemble of filaments interacting both among themselves and with the defects of the crystalline environment.\cite{Fis91,Bla94a,Bon94} In this respect, a debated topic concerns the actual occurrence of a thermodynamic phase transition between different regions in the magnetic field -- temperature ($H-T$) phase diagram of the superconducting state where dissipation processes do or do not occur.\cite{Koc89,Bis92,Str01}

The discovery of high-$T_{\textrm{c}}$ type-II superconductivity in Fe-based pnictides\cite{Kam08,Joh10,Ste11} has renewed the interest for the physics of the vortex state. These materials are widely thought to bridge the gap between high-$T_{\textrm{c}}$ cuprates and low-$T_{\textrm{c}}$ BCS-type superconductors. Besides their intermediate $T_{\textrm{c}}$ values, pnictides have indeed both high upper critical magnetic fields $H_{\textrm{c2}}$ and low values for the anisotropy parameter $\gamma$, two distinct properties of high-$T_{\textrm{c}}$ and low-$T_{\textrm{c}}$ materials, respectively. Emblematic in this respect are the materials belonging to the $122$ family, displaying $\mu_{0}H_{\textrm{c2}} \sim 50$ T while $\gamma \sim 2$ at the same time.\cite{Joh10} The role of both thermal and quantum fluctuations along the $H_{\textrm{c2}}$ vs. $T$ line resulting in precursor diamagnetism for $T \gtrsim T_{\textrm{c}}$ was discussed for Fe-based pnictides both in the $122$ and in the $1111$ families.\cite{Wel09,Mur10,Wel11,Mos11,Pra11b} At the same time, several experimental results have been reported about the thermal and quantum fluctuations in connection with the vortex pinning, the flux motion and the estimates of critical currents.\cite{Pro08,Pra10,Pra11a,Pra12,Gho12,Bos12} Nevertheless, rarely the data have been examined in terms of a critical behaviour associated with a phase transition to a glassy-like state for the vortices,\cite{Mak13} but rather in terms of thermally activated processes in the strong-disorder limit.\cite{Pra11a,Pra12} In fact, criticality is expected to be hidden by high degrees of quenched disorder in the system and, particularly, by a high efficiency of the pinning centers, as it is generally the case in pnictide materials.\cite{Esk09,Ino10,Wan10,Wan11,Pra11a,Pra12}

Here, we report on measurements of magneto-resistivity and magnetic susceptibilities, both dc and ac, performed on high-quality single crystals of Ba(Fe$_{0.9}$Co$_{0.1}$)$_{2}$As$_{2}$. Fe$_{1-x}$Co$_{x}$ is indeed among the most investigated chemical substitutions leading to superconductivity both in the $122$ and other (e.g., $1111$) families.\cite{Sef08,She10,Pra13} Our results strongly hint at the scenario of a vortex-glass phase transition. Remarkably, the results obtained from ac susceptibility are analyzed within the Havriliak-Negami framework (i.~e., a generalized formulation of the Debye relaxation) under the assumption that the correlation time $\tau$ of the vortex motion is critically-diverging in the vicinity of the glassy temperature $T_{\textrm{g}}$. Experimental results for the glassy thermodynamic line $H_{\textrm{g}}(T)$ obtained from first-harmonic magnetic ac susceptibility are in good agreement with what is deduced from magneto-resistivity. Finally, we show that the random disorder introduced by the Fe$_{1-x}$Co$_{x}$ chemical substitution is acting as quenched disorder for the vortices to a much lesser extent as compared to what was previously reported for cuprate superconductors such as, e.~g., Y$_{1-x}$Pr$_{x}$Ba$_{2}$Cu$_{3}$O$_{7-\delta}$.

\section{Experimental details and main results}

The measurements discussed in this paper, i.~e., magneto-resistivity and magnetic susceptibilities (both dc and ac) were performed on two single crystals of Ba(Fe$_{0.9}$Co$_{0.1}$)$_{2}$As$_{2}$ extracted from the same batch. The shape of both crystals is identical, namely a thin rectangular slab with the short dimension defining the crystallographic $c$-axis. Details about the crystal growth and characterization as a function of the Co content can be found in Ref.~\onlinecite{Asw11}. According to preliminary resistivity data (zero magnetic field), both samples are in conditions of almost-optimal electron doping.\cite{Asw11}

\subsection{Magneto-resistivity}\label{SectMagnetoResistivity}

\begin{figure}[t!]
\vspace{6.3cm} \includegraphics{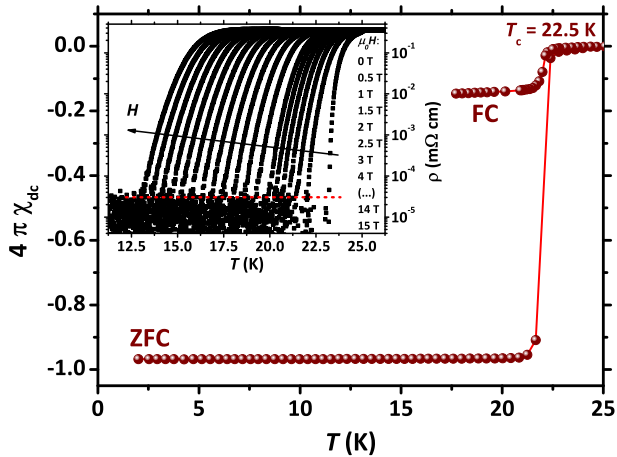} 
\caption{\label{FigChiDC_Magnetoresistivity}(Color online) Main panel: dc susceptibility vs. $T$ both in ZFC and FC conditions ($H = 5$ Oe), displaying a sharp diamagnetic onset at $T_{\textrm{c}} = 22.5 \pm 0.1$ K. Data have been corrected according to Eq.~\eqref{EqDemagnetizationDC} in order to take demagnetization effects into account. Inset: $T$ dependence of $\rho$ for different $H$ values. The red dashed line indicates the noise level. The intersections of experimental curves with this line define the temperature values $T_{\textrm{g}}^{\rho}$ for the vanishing of $\rho$.}
\end{figure}
Measurements of electrical resistivity ($\rho$) were performed as a function of $T$ by means of a standard four-probes setup. The static magnetic field $H$ was always applied at $T = 35$ K, namely far above the superconducting transition temperature $T_{\textrm{c}}$, and all the measurements were performed by slowly cooling down the crystal in field-cooled (FC) conditions. The direction of $H$ was chosen perpendicular to the main face of the crystal, namely parallel to the crystallographic $c$-axis (similarly to the case of magnetic susceptibilities, see below). Data are shown in the inset of Fig.~\ref{FigChiDC_Magnetoresistivity} up to $H = 150$ kOe.

The resistive transition of high-$T_{\textrm{c}}$ superconductors is in general sizeably broadened by the increase of $H$.\cite{Tin88,Pal88,Pal90,Saf93} This is due to dissipation effects associated with the motion of vortices, introducing an additional contribution $\rho_{\textrm{FF}}$ to the resistivity even well inside the thermodynamic superconducting phase. The subscript ``FF'' stands for ``flux flow'' (see later in Sect.~\ref{SectAnalysis}). Due to the complex interplay of several factors like the mutual repulsion among vortices, the low amount of thermal energy and the interaction with lattice defects acting like pinning centers, only in the low-$T$ region the fluxoids are not moving and turn into a ``solid'' or ``glassy'' flux phase (GFP) where irreversible magnetic phenomena occur.\cite{Yes96} On the other hand, typically pinning is not effective in the high-$T$ regime due to the increased contribution of thermal fluctuations. This latter region corresponds to the flux flow or ``liquid'' flux phase (LFP) where the fluxoids are free to move, leading in turn to dissipation. It should be remarked that the actual extension of LFP is also strongly dependent on the amount of quantum fluctuations due, e.~g., to the dimensionality of the system. Typically, quantum fluctuations are enhanced (together with the $T$ width of LFP) in systems such as Bi$_{2}$Sr$_{2}$Ca$_{2}$Cu$_{3}$O$_{10+x}$ where the very high values of $\gamma$ reduce the dimensionality of the vortex system towards the limit $D = 2$ (see Ref.~\onlinecite{Bla94a}).

From the measurements reported in the inset of Fig.~\ref{FigChiDC_Magnetoresistivity}, one has access to the temperature value $T_{\textrm{g}}^{\rho}(H)$ [or, equivalently, the field value $H_{\textrm{g}}^{\rho}(T)$] separating a true zero-resistance state where vortices are not moving from a region where, still in the presence of a thermodynamic superconducting phase, $\rho > 0$ due to the motion of vortex lines. This way, the line $T_{\textrm{g}}^{\rho}(H)$ separating liquid and solid states for the mixed Shubnikov phase is mapped out on the $H-T$ phase diagram. Accordingly, $T_{\textrm{g}}^{\rho}$ was hereby defined for every $H$ value as the $T$ value where the experimental $\rho$ vs. $T$ curves become indistinguishable from the experimental noise. In the current case of Ba(Fe$_{0.9}$Co$_{0.1}$)$_{2}$As$_{2}$ single crystals, the resistive superconducting transition is progressively shifted to lower $T$ values and broadened upon the increase of $H$. This result is in good quantitative agreement with previous results reported on similar crystals.\cite{Yam09,Wan11} It should be stressed that the $T$ extension of the extra-broadening due to the vortex motion is not as pronounced as in the case of, e.~g., $1111$ compounds.\cite{Yan08,YJo09,Lee10,Sha10} This is a clear indication of a smaller region in the $H-T$ phase diagram of Ba(Fe$_{0.9}$Co$_{0.1}$)$_{2}$As$_{2}$ where dissipative processes due to the motion of vortex lines arise. As deduced by means of magnetic ac susceptibility measurements (see the discussion below), the $T_{\textrm{g}}^{\rho}(H)$ can be addressed to as a line of thermodynamic critical points for the specific case of Ba(Fe$_{0.9}$Co$_{0.1}$)$_{2}$As$_{2}$.

\subsection{Magnetic dc susceptibility}\label{SectDCsusceptibility}

Measurements of magnetic dc susceptibility $\chi_{\textrm{dc}}$ vs. $T$ were performed by means of a Vibrating Sample Magnetometer (VSM, Quantum Design) allowing to apply polarizing static magnetic fields $H \leq 70$ kOe. The investigated single crystal has dimensions $2a = 1.7 \pm 0.025$ mm, $2b = 3.0\pm 0.025$ mm and $2c = 0.05 \pm 0.025$ mm, with $H$ applied parallel to the crystallographic $c$-axis (within an accuracy of $\pm \; 1°$), similarly to the case of the magneto-resistivity measurements discussed in Sect.~\ref{SectMagnetoResistivity}. The intrinsic susceptibility $\chi_{\textrm{dc}}^{\textrm{i}}$ is obtained from the measured one $\chi_{\textrm{dc}}^{\textrm{m}}$ as (volume units)\cite{Fio04}
\begin{equation}\label{EqDemagnetizationDC}
\frac{1}{\chi_{\textrm{dc}}^{\textrm{i}}} = \frac{1}{\chi_{\textrm{dc}}^{\textrm{m}}} - 4 \pi D_{\textrm{m}},
\end{equation}
where, for the considered geometry, the demagnetization factor is computed as $D_{\textrm{m}} \simeq 0.95$.\cite{Par04} As an example, the demagnetization-corrected $\chi_{\textrm{dc}}^{\textrm{i}}$ data for a small magnetic field ($H = 5$ Oe) are reported in the main panel of Fig.~\ref{FigChiDC_Magnetoresistivity} for zero-field cooled (ZFC) and FC conditions. Our results display a fairly sharp diamagnetic transition followed by a flattening of the ZFC data at $-1/4\pi$, indicative of a full Meissner screening from the whole volume of the sample. For $H = 5$ Oe one can deduce $T_{\textrm{c}} =  22.5 \pm 0.1$K from a double linear fitting procedure slightly above and below the diamagnetic onset.\cite{Pra11b}

\subsection{Magnetic ac susceptibility}\label{SectACsusceptibility}

Measurements of magnetic ac susceptibility $\chi_{\textrm{ac}}$ were performed by means of a Physical Properties Measurement System (PPMS, Quantum Design) allowing to apply polarizing static magnetic fields $H \leq 90$ kOe. In addition to $H$, the investigated sample is subject to a magnetic field sinusoidally-dependent on time ($t$)
\begin{equation}
H_{\textrm{ac}}(t) = H_{\textrm{ac}} \times e^{\imath \omega_{\textrm{m}} t},
\end{equation}
where $\omega_{\textrm{m}}$ represents the angular frequency and, generally, $|H_{\textrm{ac}}| \ll |H|$. In the discussed measurements, the amplitude of the alternating magnetic field $H_{\textrm{ac}}$ was always kept fixed to $1$ Oe while the frequency $\nu_{\textrm{m}} = \omega_{\textrm{m}}/2 \pi$ was swept between $10$ and $10^{4}$ Hz. Both $H$ and $H_{\textrm{ac}}$ were applied parallel to the crystallographic $c$-axis (within an accuracy of $\pm \; 1°$). The shape of the crystal is identical to the case of dc susceptibility measurements (see Sect.~\ref{SectDCsusceptibility}) and, accordingly, the value for the demagnetization factor is again $D_{\textrm{m}} \simeq 0.95$ in this geometry. The ac susceptometer allows to measure the $t$-dependent magnetization per unitary volume
\begin{equation}
M_{\textrm{ac}}(t) = M_{\textrm{ac}} \times e^{\imath \left(\omega_{\textrm{m}} t - \phi\right)}
\end{equation}
induced by $H_{\textrm{ac}}(t)$ and thus allows to resolve both the in-phase and the out-of-phase components generally arising from the $T$- and $H$-dependent phase-shift factor $\phi$.\cite{Nik94} A linear (first-harmonic) complex ac susceptibility $\chi_{\textrm{ac}}$ can be defined in terms of the Fourier transform of $M_{\textrm{ac}}(t)$ as\cite{vdB93}
\begin{eqnarray}\label{EqDefinitionComplexSusc}
\chi_{\textrm{ac}} & = & \chi_{\textrm{ac}}^{\prime} - \imath \chi_{\textrm{ac}}^{\prime\prime}\\
& = & \frac{1}{2 \pi H_{\textrm{ac}}} \times \int_{0}^{2 \pi} M_{\textrm{ac}}(t) e^{-\imath \omega_{\textrm{m}} t} d\left(\omega_{\textrm{m}} t\right).\nonumber
\end{eqnarray}
The real and imaginary components of $\chi_{\textrm{ac}}$ correspond to the in-phase and out-of-phase components of $M_{\textrm{ac}}(t)$ with respect to $H_{\textrm{ac}}(t)$, respectively. A measurement of the imaginary component of $\chi_{\textrm{ac}}$ gives direct access to the dissipation processes in the selected experimental conditions (i.~e., values of the external parameters like $T$ and $H$).\cite{vdB93}

The intrinsic susceptibility $\chi_{\textrm{ac}}^{\textrm{i}}$ is obtained from the measured one $\chi_{\textrm{ac}}^{\textrm{m}}$ by generalizing Eq.~\eqref{EqDemagnetizationDC} for the complex case, where
\begin{equation}\label{EqComplexSusceptibility}
\chi_{\textrm{ac}}^{\textrm{i,m}} = \left(\chi_{\textrm{ac}}^{\textrm{i,m}}\right)^{\prime} - \imath \; \left(\chi_{\textrm{ac}}^{\textrm{i,m}}\right)^{\prime\prime}
\end{equation}
for both quantities. Thus, one can write\cite{Dek89,Mat01} 
\begin{eqnarray}\label{EqDemagnetizationAC}
\left(\chi_{\textrm{ac}}^{\textrm{i}}\right)^{\prime} & = & \frac{\left(\chi_{\textrm{ac}}^{\textrm{m}}\right)^{\prime} - 4 \pi D_{\textrm{m}} \left\{\left[\left(\chi_{\textrm{ac}}^{\textrm{m}}\right)^{\prime}\right]^{2} + \left[\left(\chi_{\textrm{ac}}^{\textrm{m}}\right)^{\prime\prime}\right]^{2}\right\}}{\left[1 - 4 \pi D_{\textrm{m}} \left(\chi_{\textrm{ac}}^{\textrm{m}}\right)^{\prime}\right]^{2} + \left[4 \pi D_{\textrm{m}} \left(\chi_{\textrm{ac}}^{\textrm{m}}\right)^{\prime\prime}\right]^{2}}\nonumber\\
\left(\chi_{\textrm{ac}}^{\textrm{i}}\right)^{\prime\prime} & = & \frac{\left(\chi_{\textrm{ac}}^{\textrm{m}}\right)^{\prime\prime}}{\left[1 - 4 \pi D_{\textrm{m}} \left(\chi_{\textrm{ac}}^{\textrm{m}}\right)^{\prime}\right]^{2} + \left[4 \pi D_{\textrm{m}} \left(\chi_{\textrm{ac}}^{\textrm{m}}\right)^{\prime\prime}\right]^{2}}\nonumber\\
\end{eqnarray}
(volume units are assumed). In the following, only the demagnetization-corrected $\chi_{\textrm{ac}}^{\textrm{i}}$ data are considered and the superscript $i$ is dropped from now on for the sake of clarity.

\subsubsection{Isothermal $\chi_{\textrm{ac}}$ vs. $\nu_{\textrm{m}}$ scans}

Measurements of $\chi_{\textrm{ac}}$ are a powerful tool in order to investigate the low-frequency properties of the vortex motion in high-$T_{\textrm{c}}$ superconductors.\cite{Gom97} As it is recalled in detail in Sect.~\ref{SectAnalysis}, exploring the dynamic features of the penetration of the magnetic flux inside type-II superconductors gives access to the dynamics of vortices in the crossover between LFP and GFP. 

\begin{figure}[t!]
\vspace{6.3cm} \includegraphics{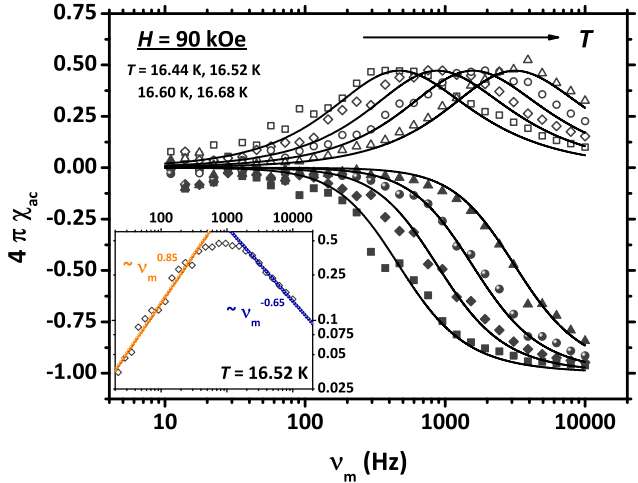} 
\caption{\label{FigFreqScans}(Color online) Main panel: real (close symbols) and imaginary (open symbols) components of complex ac susceptibility vs. frequency at representative $T$ values. Measurements have been performed at fixed values of $H = 90$ kOe and $H_{\textrm{ac}} = 1$ Oe. The continuous lines are best-fits to data according to the Havriliak-Negami framework for the magnetic relaxation ($\alpha = 1$, $\beta = 0.88$, see Sect.~\ref{SectAnalysis}). Inset: log-log plot of $\chi_{\textrm{ac}}^{\prime\prime}$ in the peak region at $T = 16.52$ K (data already presented in the main panel). The power-law exponents with absolute values $< 1$ indicate a small distribution of correlation times (see Ref.~\onlinecite{Gam90}). This is in agreement with the results of the analysis presented in Sect.~\ref{SectAnalysis}.}
\end{figure}

The most straightforward way to investigate the properties of magnetic relaxation by means of measurements of ac susceptibility is to perform scans as a function of $\nu_{\textrm{m}}$ at fixed values of both $H$ and $T$. From the shape of both real and imaginary components of the measured susceptibility one can derive the number $j$ of the main channels for the magnetic relaxation and the values (and distribution) for the relative characteristic correlation times $\tau_{j}$. Frequency scans in an external static field of $H = 90$ kOe are presented in Fig.~\ref{FigFreqScans}. The single well-defined maximum displayed in the absorptive (i.~e., imaginary) profile for each $T$ value gives a clear indication of a single main relaxation channel, namely $j = 1$ (the subscript $j$ will be dropped from now on). Accordingly, it is possible to define $\tau_{\textrm{p}} = 1/\omega_{\textrm{p}}$ as the main correlation time for the relaxation, where $\omega_{\textrm{p}} = 2 \pi \nu_{\textrm{p}}$ is the value for the angular frequency corresponding to the maximum in $\chi_{\textrm{ac}}^{\prime\prime}$. On the other hand, the absolute slope values measured by plotting $\chi_{\textrm{ac}}^{\prime\prime}$ vs. $\nu_{\textrm{m}}$ in a log-log plot (see the inset of Fig.~\ref{FigFreqScans}) are smaller than $1$, giving a qualitative indication of a small distribution of correlation times around the main value $\tau_{\textrm{p}}$ (see Ref.~\onlinecite{Gam90}). These considerations are examined in depth in Sect.~\ref{SectAnalysis} when discussing the results in terms of Cole-Cole plots and of the Havriliak-Negami framework.

Such frequency scans have been repeated at different values of $\left(H,T\right)$ in order to have access to the relative $H$ and $T$ dependencies of $\tau_{\textrm{p}}$. In the current case, we performed our measurements for $H \leq 90$ kOe always in FC conditions. The data reported in the main panel of Fig.~\ref{FigFreqScans} for $H = 90$ kOe are qualitatively representative of our results for all the other investigated $H$ values. On a quantitative level, $\tau_{\textrm{p}}$ for the observed dynamic magnetic processes increases with decreasing $T$, a feature clearly indicative of a slowing-down process. As it will be justified and discussed later in the text, such magnetic relaxation should be associated with the gradual slowing-down of the vortices in their transition from LFP to GFP.

\subsubsection{$\chi_{\textrm{ac}}$ vs. $T$ scans at fixed $\nu_{\textrm{m}}$}

\begin{figure}[t!]
\vspace{6.3cm} \includegraphics{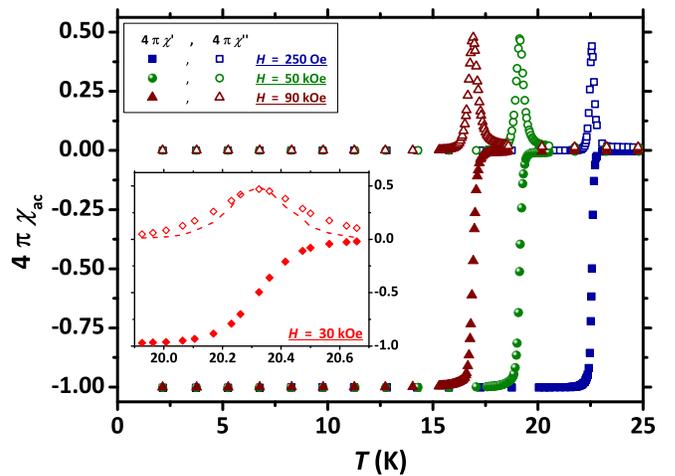} 
\caption{\label{FigSCTransitions}(Color online) Main panel: $\chi_{\textrm{ac}}$ vs. $T$ data for both real (close symbols) and imaginary (open symbols) components at $\nu_{\textrm{m}} = 9685$ Hz and at different $H$ (after correction for the demagnetization factor). Inset: zoom of data for $H = 30$ kOe around the step-like drop of the real component. The dashed line represents the derivative of $\chi_{\textrm{ac}}^{\prime}$ as a function of $T$ (arb. units).}
\end{figure}

Insights into the dynamic features of the magnetic relaxation and on the motion of vortices can also be obtained from $\chi_{\textrm{ac}}$ vs. $T$ scans at fixed $\nu_{\textrm{m}}$. Measurements were performed by slowly warming the sample after a FC procedure across $T_{\textrm{c}}$. Fig.~\ref{FigSCTransitions} shows representative results for these $T$ scans, namely a step-like diamagnetic response of the real part of $\chi_{\textrm{ac}}$, whose position is strongly dependent on the value of $H$. Remarkably, small values for the transition width $\Delta \sim 0.2$ K are detected independently on the value of $H$ (see the inset of Fig.~\ref{FigSCTransitions}). In correspondence with the sharp drop of $\chi_{\textrm{ac}}^{\prime}$, the imaginary component $\chi_{\textrm{ac}}^{\prime\prime}$ displays a single bell-shaped contribution peaked around a temperature value $T_{\textrm{p}}$ also common to the maximum in $d \chi_{\textrm{ac}}^{\prime}/d T$ (see the inset of Fig.~\ref{FigSCTransitions}).\cite{Pra11a} It should be stressed that the strong $H$-dependence is by far more pronounced than in the case of dc susceptibility. This result is in qualitative agreement with similar results obtained on high-$T_{\textrm{c}}$ superconductors\cite{Mal88,Pal90,Tin91,Pra11a,Pra12} and it confirms that measurements of $\chi_{\textrm{ac}}$ are in general not sensitive to $H_{\textrm{c}2}$ when a polarizing magnetic field $H$ is applied.\cite{Mal88} The sharp drop of $\chi_{\textrm{ac}}^{\prime}$ defines indeed the crossover between GFP and LFP of vortices separating regions inside the thermodynamic superconducting phase where dissipative processes associated with the motion of vortex lines take place or not. In fact, only deep into the GFP the superconductor is still able to effectively shield external alternating magnetic fields,\cite{Bra91,Cof92} as confirmed for the current sample by the low-$T$ saturation value $\chi_{\textrm{ac}}^{\prime} = -1/4\pi$, which is independent of $H$.\cite{Pra11a,Pra12,Mal88,Pal90,Tin91}

Due to the remarkably sharp transition width $\Delta$, the current sample is an optimal playground for investigating the intrinsic features of the superconducting phase and of the motion of the vortex lines. Details of the vortex dynamics are indeed accessed by repeating $\chi_{\textrm{ac}}$ vs. $T$ scans at a fixed $H$ and at different $\nu_{\textrm{m}}$. Following the framework of magnetic relaxation already presented for isothermal $\chi_{\textrm{ac}}$ vs. $\nu_{\textrm{m}}$ scans, it is clear that also the peaks in Fig.~\ref{FigSCTransitions} arise from the matching of $\omega_{\textrm{m}}$ to the inverse characteristic correlation time $1/\tau$.\cite{Pra11a,Pra12,Pal90,Bra92} Similarly to what is also reported in the literature in the case of cuprate superconductors (see Ref.~\onlinecite{Mal88}), a weak dependence of $T_{\textrm{p}}$ on $\omega_{\textrm{m}}$ has been detected even at the nominal $H = 0$ Oe value (in the present sample, the variation of $T_{\textrm{p}}$ at $H = 0$ Oe over the entire range of $\omega_{\textrm{m}}$ is less than $0.05$ K). Nevertheless, such effect should be considered as spurious mainly due to strong demagnetization effects possibly enhancing the sum of both $H_{\textrm{ac}}$ and any residual component of $H$ to
effective values close to the lower critical field $H_{\textrm{c1}}$. 

Summarizing, the investigation of the penetration of magnetic flux and its dynamic properties inside type-II superconductors is equivalent with the investigation of the features of the motion of vortices in their interplay with thermal energies and structural defects of the material. The peak value of $\chi_{\textrm{ac}}^{\prime\prime}$ allows to directly access the main correlation time $\tau_{\textrm{p}}$ at a fixed value of $\left(H,T\right)$ in the case of isothermal $\chi_{\textrm{ac}}$ vs. $\nu_{\textrm{m}}$ scans. Analogously, in the case of $\chi_{\textrm{ac}}$ vs. $T$ scans at fixed $\omega_{\textrm{m}}$, from the peak value of $\chi_{\textrm{ac}}^{\prime\prime}$ one defines the temperature $T_{\textrm{p}}$ where the characteristic correlation time $\tau$ matches the inverse frequency, namely $\tau = 1/\omega_{\textrm{m}}$. In the current case of one single correlation time for the investigated dynamic process, the two respective outputs $\tau_{\textrm{p}}\left(H,T\right)$ and $T_{\textrm{p}}\left(H,\tau\right)$ are equivalent.

\section{Analysis}\label{SectAnalysis}

The response of high-$T_{\textrm{c}}$ superconductors to an ac magnetic field has been investigated in detail in the literature from both theoretical and experimental points of view. In particular, the origin of the peak in the absorptive component of susceptibility has triggered great attention. Generally, all the proposed models consider the crossover between conditions where the external magnetic flux does  or does not penetrate the sample (e.~g., high- and low-$T$,  respectively) as origin of the bell-shape of $\chi_{\textrm{ac}}^{\prime\prime}$.\cite{Kes89,Lin91,Ges91,Bra91,Bra92,Cof92,vdB93} For instance, inside the LFP and under the conditions of thermally-activated flux-flow (TAFF), the peak originates once the frequency-dependent skin-depth $\delta$ of the external radiation matches the typical dimensions $d$ of the sample.\cite{Bra92} One can show that the following functional form
\begin{eqnarray}\label{EqSusceptibilityDiffusive}
4\pi\chi_{\textrm{ac}} & = & \frac{\tanh(u)}{u} - 1\\
& = & \frac{\left[\sin\left(v\right) + \sinh\left(v\right)\right] + \imath \left[\sin\left(v\right) - \sinh\left(v\right)\right]} {v\left[\cos\left(v\right) + \cosh\left(v\right)\right]} - 1\nonumber
\end{eqnarray}
holds for the magnetic ac susceptibility, where $v = d/\delta$ and $2u = v \left(1 + \imath\right)$. It is interesting to consider the explicit expression for the frequency-dependent skin-depth
\begin{equation}
\delta = \sqrt{\frac{c^{2}\rho_{\textrm{FF}}}{2\pi\omega}} \equiv \sqrt{\frac{2D_{\textrm{FF}}}{\omega}}
\end{equation}
in terms of the contribution $\rho_{\textrm{FF}}$ to the electrical resistivity coming from the liquid (flux-flow) state of vortices, $c$ being the speed of light (see also Fig.~\ref{FigChiDC_Magnetoresistivity} and the discussion in Sect.~\ref{SectMagnetoResistivity}). Equivalently, one can refer to the flux-flow diffusion constant $D_{\textrm{FF}}$ (hence the reference to ``diffusive model'' later in Fig.~\ref{FigSummarizingColeCole}). In the vicinity of the crossover between LFP and GFP for type-II superconductors, $D_{\textrm{FF}}$ is crucially influenced by the depinning of flux lines, whose typical correlation time $\tau$ satisfies the general relation
\begin{equation}\label{EqExpCorrelationTime}
\tau \sim D_{\textrm{FF}}^{-1} \sim \exp\left(\frac{U_{0}}{k_{\textrm{B}}T}\right)
\end{equation}
following the typical activated-like trend of $\rho$ in the flux-flow regime. It turns out that the maximum of $\chi_{\textrm{ac}}^{\prime\prime}$ occurs at frequencies such that $\omega_{\textrm{m}} \sim 1/\tau$.\cite{Bra92} Accordingly, the depinning energy barriers $U_{0}$ governing the $T$ dependence of $\tau$ can be derived via ac susceptibility measurements.

\subsection{Isothermal $\chi_{\textrm{ac}}$ vs. $\nu_{\textrm{m}}$ scans. Cole-Cole plots and models for $\chi_{\textrm{ac}}$}

\begin{figure}[t!]
\vspace{6.3cm} \includegraphics{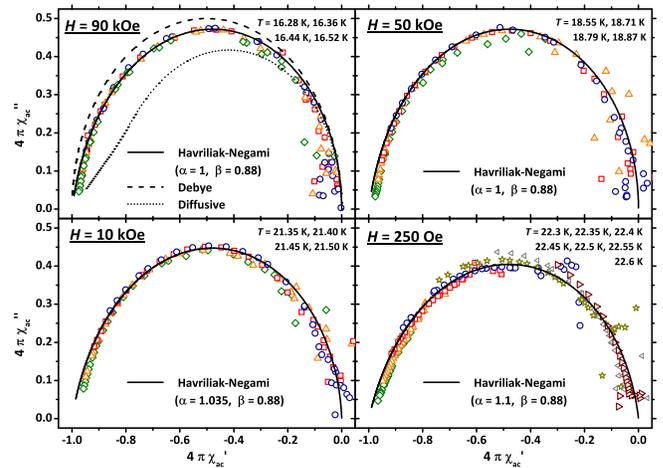} 
\caption{\label{FigSummarizingColeCole}(Color online) Cole-Cole plots for the isothermal $\chi_{\textrm{ac}}$ vs. $\nu_{\textrm{m}}$ scans at different $\left(H,T\right)$ values. Lines have been drawn according to the different relaxation models described in the text. Scans at different $T$ values collapse on single curves for each $H$ value ($H \geq 10$ kOe). The collapse is no longer well-verified at $H = 250$ Oe (see text).}
\end{figure}
A summary of our isothermal $\chi_{\textrm{ac}}$ vs. $\nu_{\textrm{m}}$ scans at different values of $H$ is reported in Fig.~\ref{FigSummarizingColeCole} (see also Fig.~\ref{FigFreqScans}). The data are presented following the canonical Cole-Cole representation, namely $\chi_{\textrm{ac}}^{\prime\prime}$ vs.  $\chi_{\textrm{ac}}^{\prime}$ with $\nu_{\textrm{m}}$ as the implicit variable.\cite{Col41,Pow51,Kub91} As it is clearly shown in the case of $H = 90$ kOe, the diffusive model for the magnetic ac susceptibility [see Eq.~\eqref{EqSusceptibilityDiffusive}] is not describing our data in the current case. Remarkably, it is apparent that our data are much more similar to a conventional Debye-like semi-circle on the Cole-Cole plot. This is quite unusual, since the Cole-Cole plot for first-harmonic $\chi_{\textrm{ac}}$ of high-$T_{\textrm{c}}$ superconductors is typically much more distorted, also according to the diffusive model discussed above.\cite{Lin91,Emm91,Her97} However, some slight degree of distortion from the ideal case is detected, possibly arising from a narrow distribution of correlation times, in agreement with the qualitative considerations in the inset of Fig.~\ref{FigFreqScans}. A phenomenological way to take this distribution into account is to consider the so-called Havriliak-Negami relaxation function within the Casimir-Du Pr\'e approximation\cite{Kub91,Hav66,Cav84,Mor01}
\begin{equation}\label{EqHavriliakNegami}
\chi_{\textrm{ac}} = \chi_{\infty} + \frac{\chi_{0} - \chi_{\infty}}{\left[1 - \left(\imath \omega_{\textrm{m}}\tau\right)^{\alpha}\right]^{\beta}}.
\end{equation}
Here, the two phenomenological exponents $\alpha$ and $\beta$ allow to properly tune the height of the maximum  of $\chi_{\textrm{ac}}^{\prime\prime}$ and the skewness in the Cole-Cole plot, respectively. For $\alpha = 1$ one typically refers to the Davidson-Cole formalism, while in the case $\alpha = 1$ and $\beta = 1$ one deals with the standard Debye relaxation. From a comparison with the main panel of Fig.~\ref{FigFreqScans} one has
\begin{equation}\label{EqCasimirDuPre}
\chi_{0} = 0 \qquad \textrm{and} \qquad \chi_{\infty} = - \frac{1}{4\pi},
\end{equation}
assuming that the considered $T$ range is well below the thermodynamic $T_{\textrm{c}}(H)$.

As already anticipated in the main panel of Fig.~\ref{FigFreqScans}, the Davidson-Cole framework correctly fits our data for high-$H$ values ($\beta = 0.88$). It is shown in Fig.~\ref{FigSummarizingColeCole} that the same result well applies for $H > 10$ kOe, while for $H \leq 10$ kOe some degree of skewness should be introduced via the exponent $\alpha$ (this exponent increasing with decreasing $H$). It should also be remarked that our data at different $T$ values always collapse onto a well-defined curve with the exception of the data at $H = 250$ Oe. The reason is that the crossover between LFP and GFP comes closer and closer to $T_{\textrm{c}}$ at low-$H$ so that the diamagnetic susceptibility is not fully saturated yet for all the investigated $T$ values. Namely, the condition $\chi_{\infty} = -1/4\pi$ expressed in Eq.~\eqref{EqCasimirDuPre} is no longer holding at too low-$H$ values.

\subsection{Critical divergence of correlation time}

\begin{figure}[t!]
\vspace{6.3cm} \includegraphics{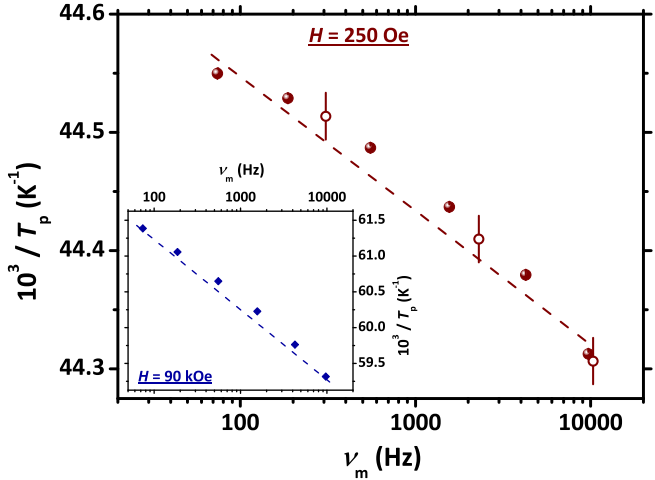}
\caption{\label{FigFreqDep}(Color online) Main panel: $1/T_{\textrm{p}}$ as a function of $\nu_{\textrm{m}}$ (full symbols). Open symbols refer to the estimates from isothermal $\chi_{\textrm{ac}}^{\prime\prime}$ vs.$\nu_{\textrm{m}}$ curves (see Sect.~\ref{SectACsusceptibility}). The dashed line shows the expected logarithmic trend typical of thermally-activated processes according to Eq.~\eqref{EqLogDep}. Data are taken at $H = 250$ Oe (data for $H = 90$ kOe are plotted in the inset with the same meaning of symbols).}
\end{figure}
Further insight into the $T$ dependence of the correlation time $\tau$ can be achieved from the analysis of $\chi_{\textrm{ac}}$ vs. $T$ scans (see Sect.~\ref{SectACsusceptibility}). First, let us consider only the dependence of $T_{\textrm{p}}$ on $\omega_{\textrm{m}}$. If the degree of quenched disorder in the system is sizeable and, correspondingly, the density of pinning centers is high, thermally-activated processes are the leading mechanisms in driving the dynamic relaxation of the fluxoids. This results in a logarithmic dependence\cite{Mal88,Pra11a,Pra12,Lee13}
\begin{equation}\label{EqLogDep}
\frac{1}{T_{\textrm{p}}\left(\omega_{m}\right)} = -\frac{k_{\textrm{B}}}{U_{0}} \times \ln\left(\frac{\omega_{m}}{\omega_{0}}\right)
\end{equation}
governed by the depinning energy barrier $U_{0}$ already introduced in Eq.~\eqref{EqExpCorrelationTime}. Here, $k_{\textrm{B}} = 1.38 \times 10^{-16}$ erg/K represents the Boltzmann constant. However, this expression is not a good choice in order to describe our data for the current single crystals of Ba(Fe$_{0.9}$Co$_{0.1}$)$_{2}$As$_{2}$, as it is shown in Fig.~\ref{FigFreqDep} for the two limiting cases $H = 250$ Oe and $H = 90$ kOe. The deviation from the dashed line, representing the expected trend according to Eq.~\eqref{EqLogDep}, is particularly evident for $H = 250$ Oe. The effect is still present at $H = 90$ kOe, even if less pronounced (see the inset of Fig.~\ref{FigFreqDep}).

A different approach in order to describe our experimental data comes from the theory of phase transitions and of critical phenomena.\cite{Sta71} By focussing on the case of a clean superconducting system in the limit of weak disorder, opposite to what has been considered above, the lower concentration of pinning centers makes the interaction among vortices drive the magnetic relaxation. This allows the detection of a scaling critical behaviour associated with a thermodynamic phase transition for the flux lines between LFP and GFP. The fingerprint of criticality is the appearance of power-law functional forms describing the trend of physical quantities as a function of some reduced variable quantifying the distance of the system from the critical point.\cite{Sta71} With reference to the model discussed in Refs.~\onlinecite{Fis91} and \onlinecite{Gam90}, one can assume that the transition of vortices from LFP to GFP occurs at a critical value $T_{\textrm{g}}$ dependent on the value of $H$. Values of $T_{\textrm{g}}(H)$ then define the vortex-glass melting line, equivalently referred to as $H_{\textrm{g}}(T)$. From dynamic scaling considerations the following relation can be derived for the reduced temperature $\varepsilon_{\textrm{g}}(\omega_{\textrm{m}})$ (see Refs.~\onlinecite{Gam90,Sag90,Fis91})
\begin{equation}\label{EqScaling}
\varepsilon_{\textrm{g}}(\omega_{\textrm{m}}) \equiv \frac{T_{\textrm{p}}(\omega_{\textrm{m}})-T_{\textrm{g}}^{\chi}} {T_{\textrm{g}}^{\chi}} = \left(\frac{\omega_{\textrm{m}}}{\omega_{0}}\right)^{1/s},
\end{equation}
where the superscript $\chi$ stresses that the estimate was performed starting from susceptibility data. The parameter $\omega_{0}$ in this expression is a microscopic characteristic angular frequency with typical values in the THz range.\cite{Gam90} For the critical exponent the relation $s = \left[\nu\left(z+2-D\right)\right]$ holds, where the dimensionality of the vortex system is indicated by $D$, while $z$ and $\nu$ are the dynamic and the static critical exponents for the vortex glass, respectively. It should be remarked that this approach is the same as the one employed in the case of magnetic spin glasses even if, in that case, the relation holding for the critical exponent is $s_{\textrm{SG}} = \nu z$.\cite{Cha11,Hoh77}

\begin{figure}[t!]
\vspace{6.3cm} \includegraphics{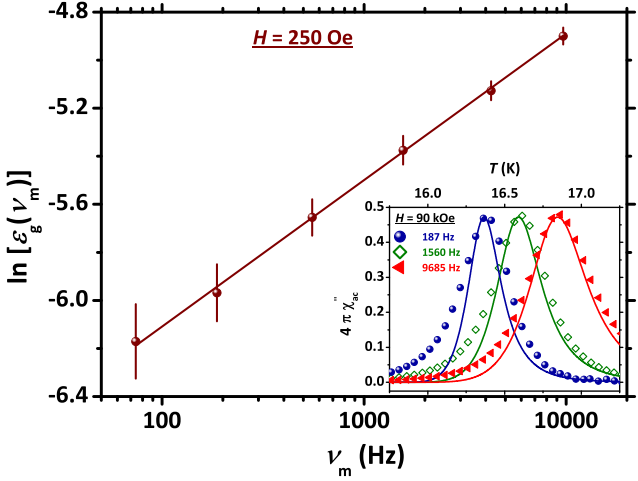}  
\caption{\label{FigFreqDepScaling}(Color online) Main panel: $\ln\left(\varepsilon_{\textrm{g}}\right)$ vs. $\nu_{\textrm{m}}$ semi-log plot at $H = 250$ Oe (see text). The dashed line is a fit to our experimental data according to Eq.~\eqref{EqScaling}. Inset: experimental data for the $T$ dependence of the complex component of the ac susceptibility measured at $H = 90$ kOe and at different frequencies. The continuous lines are the result of a simultaneous fitting procedure of all the curves according to Eq.~\eqref{EqHavriliakNegami} (the values for $\alpha$ and $\beta$ being reported in Fig.~\ref{FigSummarizingColeCole}) and Eq.~\eqref{EqCritCorrelationTime}.}
\end{figure}
Our data have been examined according to this model and, in particular, to Eq.~\eqref{EqScaling}. A representative $\varepsilon_{\textrm{g}}$ vs. $\nu_{\textrm{m}}$ curve has been plotted in Fig.~\ref{FigFreqDepScaling} after a proper selection of $T_{\textrm{g}}^{\chi}$ (for $H = 250$ Oe). The accuracy associated with the assessment of the $T_{\textrm{g}}^{\chi}$ is $\sim 0.02$ K. This high precision in the estimate of $T_{\textrm{g}}^{\chi}$ allows us to strongly reduce the mutual statistical dependence among the parameters $s$ and $\omega_{0}$ when fitting the experimental data to Eq.~\eqref{EqScaling}. Due to the tiny variation of the absolute values of $T_{\textrm{p}}$ upon varying $\omega_{\textrm{m}}$, the fitting procedure gives access to physical quantities \emph{along} $H_{\textrm{g}}^{\chi}(T)$. It is clear that our results for the current single crystal unambiguously reveal a scaling behaviour rather than a logarithmic one.

As further check of the scenario outlined above, we considered the functional form for the $\chi_{\textrm{ac}}$ susceptibility derived within the Havriliak-Negami relaxation framework [see Eq.~\eqref{EqHavriliakNegami}] and tried to reproduce the $T$ dependence of $\chi_{\textrm{ac}}^{\prime\prime}$ at fixed $\nu_{\textrm{m}}$ by describing the $T$ dependence of the correlation time $\tau$. In particular, an expression for critically slowing-down processes can be written as
\begin{equation}\label{EqCritCorrelationTime}
\tau(T) = \tau_{0} \times \left(\frac{T-T_{\textrm{g}}^{\chi}}{T_{\textrm{g}}^{\chi}}\right)^{-s^{\prime}} = \tau_{0} \times \varepsilon_{\textrm{g}}^{-s^{\prime}}.
\end{equation}
The substitution of Eq.~\eqref{EqCritCorrelationTime} into Eq.~\eqref{EqHavriliakNegami} gives rise to the solid curves displayed in the inset of Fig.~\ref{FigFreqDepScaling}, where our data are reported at the fixed representative field value $H = 90$ kOe and for different values of $\nu_{\textrm{m}}$. The agreement with our experimental data is rather good. In general, the quality of the simulation is increasing with increasing $\nu_{\textrm{m}}$. At the same time, the quality of fitting is worse on the low-$T$ side of the peaks, while the peak region and the high-$T$ side are well reproduced. The low-$T$ discrepancy may arise from non-linear effects likely appearing extremely close to $T_{\textrm{g}}^{\chi}$ and giving a sizeable extra-contribution to the absorption.\cite{Ish90,Lam90,vdB93,Pol00,Ade04,Pol08} However, it should be stressed that the position of the peaks is always well-reproduced by the simulated curve. Moreover, and most importantly, the curves are not individually fitted but the considered set of fitting parameters is fixed once $H$ is fixed. In particular, both $\tau_{0} = 1/\omega_{0}$ and $T_{\textrm{g}}^{\chi}$ were chosen after fitting the data presented in the main panel of Fig.~\ref{FigFreqDepScaling} according to Eq.~\eqref{EqScaling} (the actual values for the resulting parameters are presented and discussed in Sect.~\ref{SectDiscussion}). The critical exponent $s^{\prime}$ was the only parameter which was left as a free variable, and which is allowed to be different from $s$ obtained from Eq.~\eqref{EqScaling}. Remarkably, one finds $s(H) = s^{\prime}(H)$ as output for every $H$ value. Accordingly, $s^{\prime} = \left[\nu\left(z+2-D\right)\right]$.
\begin{figure}[t!]
\vspace{6.3cm} \includegraphics{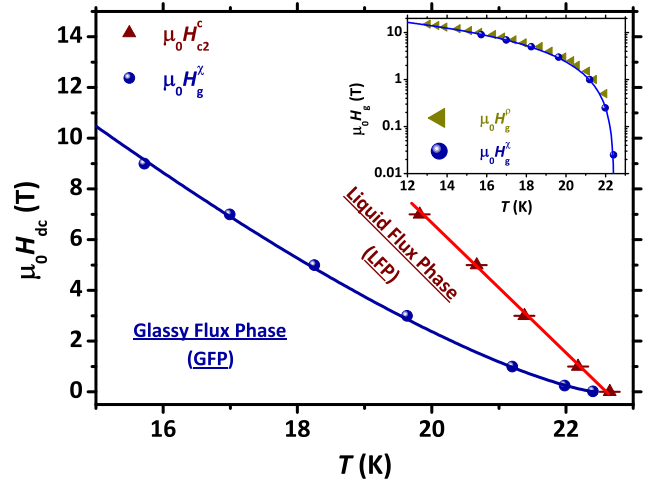} 
\caption{\label{FigPhDiag}(Color online) Main panel: phase diagram for the superconducting state of Ba(Fe$_{0.9}$Co$_{0.1}$)$_{2}$As$_{2}$. The solid blue line is a fit to the $H_{\textrm{g}}^{\chi}$ data according to the power-law function discussed in the text, while the solid red line is a linear fit to the $H_{\textrm{c2}}^{\textrm{c}}$ data. $H_{\textrm{g}}^{\chi}$ data are reported also in the inset in a semi-log plot. Estimates of $H_{\textrm{g}}^{\rho}$ obtained from the magneto-resistivity data discussed in Sect.~\ref{SectMagnetoResistivity} are also reported in the inset.}
\end{figure}

\section{Discussion. $H$ -- $T$ Phase Diagram}\label{SectDiscussion}

\begin{figure}[t!]
\vspace{6.3cm} \includegraphics{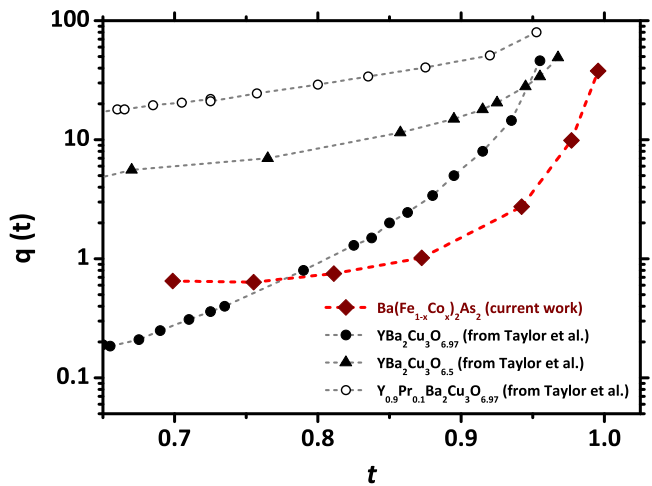} 
\caption{\label{FigQuantumFluct}(Color online) Main panel: $T$ dependence of the parameter $q$ quantifying the amount of quantum fluctuations along the melting line according to Eq.~\eqref{EqQuantFluct}. Our data are compared with results obtained for YBa$_{2}$Cu$_{3}$O$_{7-\delta}$ and Y$_{1-x}$Pr$_{x}$Ba$_{2}$Cu$_{3}$O$_{6.97}$ (data are reproduced from Ref.~\onlinecite{Tay07}). The dashed lines are a guides-to-the-eye.}
\end{figure}
As discussed in the previous Sections, the most important output of our measurements is the estimate of the critical temperature $T_{\textrm{g}}(H)$ [or, equivalently, $H_{\textrm{g}}(T)$] for the liquid-glass transition of vortices. The $H_{\textrm{g}}^{\chi}(T)$ values obtained from ac susceptibility measurements are plotted in Fig.~\ref{FigPhDiag}. Our data for $H_{\textrm{g}}^{\chi}(T)$ agree well with previous results obtained for Ba(Fe$_{1-x}$Co$_{x}$)$_{2}$As$_{2}$ via dc magnetometry.\cite{Ino10} The relative $T$ trend can be fitted by a power-law like function $H_{\textrm{g}}^{\chi} \propto \left[T_{\textrm{c}}(0) - T_{\textrm{g}}^{\chi}\right]^{\beta}$ where $\beta = 1.30 \pm 0.05$ (see the solid blue line in Fig.~\ref{FigPhDiag}), in excellent agreement with the expected result $\beta = 4/3$ for the vortex-glass model for $D = 3$ and with previous reports, e.~g., on YBa$_{2}$Cu$_{3}$O$_{7-\delta}$ and other cuprate superconductors.\cite{Fis91,Gam91,Bla94a} A similar value for the exponent $\beta$ was reported recently in hole-doped $122$ systems.\cite{Mak13} It should be remarked that the value $\beta = 4/3$ is also in agreement with the typical findings for the so-called De Almeida - Thouless line in magnetic spin glasses.\cite{Cha11} 

In order to further stress the robustness of our analysis of ac susceptibility measurements, we plot a comparison with the $H_{\textrm{g}}^{\rho}(T)$ values obtained from magneto-resistivity (see Sect.~\ref{SectMagnetoResistivity}) in the inset of Fig.~\ref{FigPhDiag}, showing a very good agreement among the two sets of data. The results for $T_{\textrm{g}}^{\chi}\left(H\right)$ are complemented in Fig.~\ref{FigPhDiag} by the $T$ dependence of $H_{\textrm{c2}}^{\textrm{c}}$ derived from measurements of dc magnetization at different $H$ values (see Sect.~\ref{SectDCsusceptibility}). A slope value $-2.50 \pm 0.05$ T/K is derived for this latter quantity, in perfect agreement with previous reports on Ba(Fe$_{1-x}$Co$_{x}$)$_{2}$As$_{2}$.\cite{Yam09} Concerning the other parameters obtained from the analysis of our ac susceptibility data, the frequency $\nu_{0}$ shows values $\sim 10^{13} - 10^{15}$ Hz, slightly higher than other reports in the literature.\cite{Gam90} On the other hand, the critical exponent $s \sim 4 - 7$ is in close agreement with previous reports on cuprate compounds like YBa$_{2}$Cu$_{3}$O$_{7}$ and Bi$_{2}$Sr$_{2}$CaCu$_{2}$O$_{8+\delta}$, where values $\sim 6 - 8$ are reported.\cite{Koc89,Gam90,Gam91,Saf92}

The detection of a low level of quenched disorder in the system is crucial in order to check for the reliability of the scenario depicted above. In order to get further information about this point, we quantified the contribution of quantum fluctuations in comparison to thermal fluctuations along the melting line. It should
be stressed that, as argued in Ref.~\onlinecite{Tay07}, the amount of quantum fluctuations is generally enhanced by the quenched disorder so that, accordingly, the detection of a low contribution of quantum fluctuations can be considered as a strong evidence of a low degree of disorder. The estimate can be obtained by means of
the function (see Ref.~\onlinecite{Tay07})
\begin{equation}\label{EqQuantFluct}
q(t) = \Pi_{0}(t) \times \sqrt{\frac{\beta_{\textrm{th}}}{Gi\left[H_{\textrm{g}}(t)\right]}}.
\end{equation}
Here, $t \equiv T/T_{\textrm{c}}(0)$ and $\beta_{\textrm{th}} = 5.6$. For the field-dependent Ginzburg number evaluated along the melting line the expression\cite{Tay07}
\begin{equation}
Gi\left[H_{\textrm{g}}(t)\right] = \sqrt[3]{\frac{\pi H_{\textrm{g}}^{2}(t)}{\Phi_{0} H_{\textrm{c}2}^{\textrm{c}}(0)} \times \left[\frac{k_{\textrm{B}}T_{\textrm{c}}(0) 8 \pi^{2} \lambda_{\textrm{ab}}^{2}(0) \gamma}{\Phi_{0}^{2}}\right]^{2}}
\end{equation}
holds, where $H_{\textrm{c}2}^{\textrm{c}}(0) \simeq 5.5 \times 10^{5}$ Oe at zero $T$\cite{Yam09} and $\lambda_{\textrm{ab}}(0)$ is the in-plane penetration depth at zero $T$. $\mu^{+}$-spin spectroscopy measurements on Ba(Fe$_{1-x}$Co$_{x}$)$_{2}$As$_{2}$ across the phase diagram yield the value $\lambda_{\textrm{ab}}(0) \simeq 180$ nm for the sample whose $T_{\textrm{c}}(0)$ value matches our findings.\cite{Wil10} The function $\Pi_{0}(t)$ can be expressed in terms of the Lindemann number $c_{\textrm{L}} = \pi^{-1/2} \left\{Gi\left[H_{\textrm{g}}(t)\right]\right\}^{1/8\beta}$ (see Ref.~\onlinecite{Tay07b}) as
\begin{widetext}
\begin{equation}
\Pi_{0}(t) = \left\{\frac{t^{2}}{1-t} \times \frac{Gi\left[H_{\textrm{g}}(t)\right]} {4 \beta_{\textrm{th}} c_{\textrm{L}}^{2}} \times \left[-1 + \left(-1 + \sqrt{\frac{4 H_{\textrm{c}2}^{\textrm{c}}(0)} {H_{\textrm{g}}(t)} \frac{\beta_{\textrm{th}} c_{\textrm{L}}^{4}} {Gi\left[H_{\textrm{g}}(t)\right]} \left(\frac{1-t}{t}\right)^{2}}\right)^{2}\right]\right\} - c_{\textrm{L}}^{2}.
\end{equation}
\end{widetext}

The resulting $T$ dependence of $q$ is presented in Fig.~\ref{FigQuantumFluct}. Remarkably, the degree of quantum fluctuations in Ba(Fe$_{0.9}$Co$_{0.1}$)$_{2}$As$_{2}$ is relatively low, as it is deduced from a comparison with, e. g., Y$_{1-x}$Pr$_{x}$Ba$_{2}$Cu$_{3}$O$_{7-\delta}$ (see Fig.~$3$ in Ref.~\onlinecite{Tay07}. The same data have been reproduced in Fig.~\ref{FigQuantumFluct} in a comparable range of substitution values $x \leq 0.1$). In the case of Y$_{1-x}$Pr$_{x}$Ba$_{2}$Cu$_{3}$O$_{7-\delta}$, a non-negligible substitution value of $x = 0.1$ results in sizeably increased $q$ values if compared to YBa$_{2}$Cu$_{3}$O$_{7-\delta}$ in the clean limit.\cite{Tay07} By quantitatively comparing the results for Ba(Fe$_{1-x}$Co$_{x}$)$_{2}$As$_{2}$ with those for Y$_{1-x}$Pr$_{x}$Ba$_{2}$Cu$_{3}$O$_{7-\delta}$ one realizes that in our case an identical amount of chemical Fe$_{1-x}$Co$_{x}$ substitution ($x = 0.1$) does not substantially increase the amount of quantum fluctuations (see Fig.~\ref{FigQuantumFluct}). Thus, it can be concluded that the random disorder introduced by Fe$_{1-x}$Co$_{x}$ doping does not introduce any appreciable quenched disorder for the vortex phase in Ba(Fe$_{0.9}$Co$_{0.1}$)$_{2}$As$_{2}$. These favourable conditions give strong hints towards the robustness of the analysis depicted above.

\section{Conclusions}

In summary, the high quality of the investigated single crystals of superconducting Ba(Fe$_{0.9}$Co$_{0.1}$)$_{2}$As$_{2}$ allows us to evidence a critical behaviour associated with the phase transition of vortex lines from liquid to glassy states by means of magneto-resistivity and magnetic ac susceptibility measurements. Remarkably, the diffusive model is not reproducing the frequency-dependence of both the real and imaginary components of the ac magnetic susceptibility. Instead, a generalized Debye-relaxation model (Havriliak-Negami framework) is employed in order to correctly reproduce the experimental data. Both frequency- and temperature-dependent scans are analyzed according to this approach. The critically-divergent temperature-dependence of the correlation time for vortices confirms that the crossover between the liquid and the glassy regions of flux lines should be considered as a phase transition. Accordingly, the values for the critical glass temperature and the critical glass exponent are derived. At variance with what was previously reported for cuprate superconductors such as, e.~g., Y$_{1-x}$Pr$_{x}$Ba$_{2}$Cu$_{3}$O$_{7-\delta}$, the random disorder introduced by the Fe$_{1-x}$Co$_{x}$ chemical substitution is found to introduce a relatively low degree of quenched disorder for the vortex phase. Overall, our results open the way to a more detailed investigation of the physics of vortices across the phase diagram of Ba(Fe$_{1-x}$Co$_{x}$)$_{2}$As$_{2}$. The evolution of the degree of glassiness and its possible interplay with magnetic fluctuations arising from the progressively enhanced nesting of the Fermi surface with reducing $x$, in particular, could be an extremely peculiar study in the future.

\section*{Acknowledgements}

Stimulating discussions with P. Carretta and S. Singh are gratefully acknowledged. The authors thank M. Deutschmann, S. Pichl and S. Gaß for technical support. G. P. acknowledges support by the Leibniz-Deutscher Akademischer Austausch Dienst (DAAD) and Alexander von Humboldt Post-Doc Fellowship Programs. S. W. acknowledges support by the Deutsche Forschungsgemeinschaft (DFG) under the Emmy-Noether program (Grant No. WU595/3-1) and the BMBF for support in the framework of the ERA.Net
RUS project. Work was supported by DFG through the Priority Programme SPP1458 (Grants No. BE1749/13 and BU887/15-1).




\begin{references}
\bibitem{Fis91} D. S. Fisher, M. P. A. Fisher, D. A. Huse, {\it Phys. Rev. B} {\bf 43}, 130 (1991)
\bibitem{Bla94a} G. Blatter, M. V. Feigel'man, V. B. Geshkenbein, A. I. Larkin, V. M. Vinokur, {\it Rev. Mod. Phys.} {\bf 66}, 1125 (1994)
\bibitem{Bon94} {\it The Vortex State} (edited by N. Bontemps, Y. Bruynseraede, G, Deutscher, A. Kapitulnik), Kluwer Academic Publishers (1994)
\bibitem{Koc89} R. H. Koch, V. Foglietti, W. J. Gallagher, G. Koren, A. Gupta, M. P. A. Fisher, {\it Phys. Rev. Lett.} {\bf 63}, 1511 (1989)
\bibitem{Bis92} D. J. Bishop, P. L. Gammel, D. A. Huse, C. A. Murray, {\it Science} {\bf 255}, 165 (1992)
\bibitem{Str01} D. R. Strachan, M. C. Sullivan, P. Fournier, S. P. Pai, T. Venkatesan, C. J. Lobb, {\it Phys. Rev. Lett.} {\bf 87}, 067007 (2001)
\bibitem{Kam08} Y. Kamihara, T. Watanabe, M. Hirano, H. Hosono, {\it J. Am. Chem. Soc.} {\bf 130}, 3296 (2008)
\bibitem{Joh10} D. C. Johnston, {\it Adv. Phys.} {\bf 59}, 803 (2010)
\bibitem{Ste11} G. R. Stewart, {\it Rev. Mod. Phys.} {\bf 83}, 1589 (2011)
\bibitem{Wel09} U. Welp, R. Xie, A. E. Koshelev, W. K. Kwok, H. Q. Luo, Z. S. Wang, G. Mu, H. H. Wen, {\it Phys. Rev. B} {\bf 79}, 094505 (2009)
\bibitem{Mur10} J. M. Murray, Z. Tesanovic, {\it Phys. Rev. Lett.} {\bf 105}, 037006 (2010)
\bibitem{Wel11} U. Welp, C. Chaparro, A. E. Koshelev, W. K. Kwok, A. Rydh, N. D. Zhigadlo, J. Karpinski, S. Weyeneth, {\it Phys. Rev. B} {\bf 83}, 100513 (2011)
\bibitem{Mos11} J. Mosqueira, J. D. Dancausa, F. Vidal, S. Salem-Sugui, A. D. Alvarenga, H.-Q. Luo, Z.-S. Wang, H.-H. Wen, {\it Phys. Rev. B} {\bf 83}, 094519 (2011)
\bibitem{Pra11b} G. Prando, A. Lascialfari, A. Rigamonti, L. Roman\'o, S. Sanna, M. Putti, M. Tropeano, {\it Phys. Rev. B} {\bf 84}, 064507 (2011)
\bibitem{Pro08} R. Prozorov, N. Ni, M. A. Tanatar, V. G. Kogan, R. T. Gordon, C. Martin, E. C. Blomberg, P. Prommapan, J. Q. Yan, S. L. Bud'ko, P. C. Canfield, {\it Phys. Rev. B} {\bf 78}, 224506 (2008)
\bibitem{Pra10} A. K. Pramanik, L. Harnagea, S. Singh, S. Aswartham, G. Behr, S. Wurmehl, C. Hess, R. Klingeler, B. B\"uchner, {\it Phys. Rev. B} {\bf 82}, 014503 (2010)
\bibitem{Pra11a} G. Prando, P. Carretta, R. De Renzi, S. Sanna, A. Palenzona, M. Putti, M. Tropeano, {\it Phys. Rev. B} {\bf 83}, 174514 (2011)
\bibitem{Pra12} G. Prando, P. Carretta, R. De Renzi, S. Sanna, H.-J. Grafe, S. Wurmehl, B. B\"uchner, {\it Phys. Rev. B} {\bf 85}, 144522 (2012)
\bibitem{Gho12} S. R. Ghorbani, X. L. Wang, M. Shahbazi, S. X. Dou, K. Y. Choi, C. T. Lin, {\it Appl. Phys. Lett.} {\bf 100}, 072603 (2012)
\bibitem{Bos12} L. Bossoni, P. Carretta, A. Thaler, P. C. Canfield, {\it Phys. Rev. B} {\bf 85}, 104525 (2012); L. Bossoni, P. Carretta, M. Horvatic, M. Corti, A. Thaler, P. C. Canfield, {\it Europhys. Lett.} {\bf 102}, 17005 (2013)
\bibitem{Mak13} H. K. Mak, P. Burger, L. Cevey, T. Wolf, C. Meingast, R. Lortz, {\it Phys. Rev. B} {\bf 87}, 214523 (2013)
\bibitem{Esk09} M. R. Eskildsen, L. Y. Vinnikov, I. S. Veshchunov, T. M. Artemova, T. D. Blasius, J. M. Densmore, C. D. Dewhurst, N. Ni, A. Kreyssig, S. L. Bud'ko, P. C. Canfield, A. I. Goldman, {\it Physica C} {\bf 469}, 529 (2009)
\bibitem{Ino10} D. S. Inosov, T. Shapoval, V. Neu, U. Wolff, J. S. White, S. Haindl, J. T. Park, D. L. Sun, C. T. Lin, E. M. Forgan, M. S. Viazovska, J. H. Kim, M. Laver, K. Nenkov, O. Khvostikova, S. K\"uhnemann, V. Hinkov, {\it Phys. Rev. B} {\bf 81}, 014513 (2010)
\bibitem{Wan10} X.-L. Wang, S. R. Ghorbani, S.-I. Lee, S. X. Dou, C. T. Lin, T. H. Johansen, K.-H. M\"uller, Z. X. Cheng, G. Peleckis, M. Shabazi, A. J. Qviller, V. V. Yurchenko, G. L. Sun, D. L. Sun, {\it Phys. Rev. B} {\bf 82}, 024525 (2010)
\bibitem{Wan11} L. M. Wang, U.-C. Sou, H. C. Yang, L. J. Chang, C.-M. Cheng, K.-D. Tsuei, Y. Su, T. Wolf, P. Adelmann, {\it Phys. Rev. B} {\bf 83}, 134506 (2011)
\bibitem{Sef08} A. S. Sefat, R. Jin, M. A. McGuire, B. C. Sales, D. J. Singh, D. Mandrus, {\it Phys. Rev. Lett.} {\bf 101}, 117004 (2008)
\bibitem{She10} B. Shen, P. Cheng, Z. Wang, L. Fang, C. Ren, L. Shan, H.-H. Wen, {\it Phys. Rev. B} {\bf 81}, 014503 (2010)
\bibitem{Pra13} G. Prando, O. Vakaliuk, S. Sanna, G. Lamura, T. Shiroka, P. Bonf\'a, P. Carretta, R. De Renzi, H.-H. Klauss, C. G. F. Blum, S. Wurmehl, C. Hess, and B. B\"uchner, {\it Phys. Rev. B} {\bf 87}, 174519 (2013)
\bibitem{Asw11} S. Aswartham, C. Nacke, G. Friemel, N. Leps, S. Wurmehl, N. Wizent, C. Hess, R. Klingeler, G. Behr, B\"uchner, {\it J. Cryst. Growth} {\bf 314}, 341 (2011)
\bibitem{Tin88} M. Tinkham, {\it Phys. Rev. Lett.} {\bf 61}, 1658 (1988)
\bibitem{Pal88} T. T. M. Palstra, B. Batlogg, L. F. Schneemeyer, J. V. Waszczak, {\it Phys. Rev. Lett.} {\bf 61}, 1662 (1988)
\bibitem{Pal90} T. T. M. Palstra, B. Batlogg, R. B. van Dover, L. F. Schneemeyer, J. V. Waszczak, {\it Phys. Rev. B} {\bf 41}, 6621 (1990)
\bibitem{Saf93} H. Safar, P. L. Gammel, D. A. Huse, D. J. Bishop, W. C. Lee, J. Giapintzakis, D. M. Ginsberg, {\it Phys. Rev. Lett.} {\bf 70}, 3800 (1993)
\bibitem{Yes96} Y. Yeshurun, A. P. Malozemoff, A. Shaulov, {\it Rev. Mod. Phys.} {\bf 68}, 911 (1996)
\bibitem{Yam09} A. Yamamoto, J. Jaroszynski, C. Tarantini, L. Balicas, J. Jiang, A. Gurevich, D. C. Larbalestier, R. Jin, A. S. Sefat, M. A. McGuire, B. C. Sales, D. K. Christen, D. Mandrus, {\it Appl. Phys. Lett.} {94}, 062511 (2009)
\bibitem{Yan08} H. Yang, C. Ren, L. Shan, H.-H. Wen, {\it Phys. Rev. B} {\bf 78}, 092504 (2008)
\bibitem{YJo09} Y. J. Jo, J. Jaroszynski, A. Yamamoto, A. Gurevich, S. C. Riggs, G. S. Boebinger, D. Larbalestier, H. H. Wen, N. D. Zhigadlo, S. Katrych, Z. Bukowski, J. Karpinski, R. H. Liu, H. Chen, X. H. Chen, L. Balicas, {\it Physica C} {\bf 469}, 566 (2009)
\bibitem{Lee10} H.-S. Lee, M. Bartkowiak, J. S. Kim, H.-J. Lee, {\it Phys. Rev. B} {\bf 82}, 104523 (2010)
\bibitem{Sha10} M. Shahbazi, X. L. Wang, C. Shekhar, O. N. Srivastava, S. X. Dou, {\it Supercond. Sci. Technol.} {\bf 23}, 105008 (2010)
\bibitem{Fio04} F. Fiorillo, {\it Measurement and Characterization of Magnetic Materials}, Elsevier Academic Press (2004), pages 8-16 
\bibitem{Par04} E. Pardo, D.-X. Chen, A. Sanchez, {\it J. Appl. Phys.} {\bf 96}, 5365 (2004)
\bibitem{Nik94} M. Nikolo, {\it Am. J. Phys.} {\bf 63}, 57 (1994)
\bibitem{vdB93} C. J. van der Beek, V. B. Geshkenbein, V. M. Vinokour, {\it Phys. Rev. B} {\bf 48}, 3393 (1993)
\bibitem{Dek89} C. Dekker, A. F. M. Arts, H. W. de Wijn, A. J. van Duyneveldt, J. A. Mydosh, {\it Phys. Rev. B} {\bf 40}, 11243 (1989)
\bibitem{Mat01} K. Matsuhira, Y. Hinatsu, T. Sakakibara, {\it J. Phys.: Cond. Matt.} {\bf 13}, L737 (2001)
\bibitem{Gom97} F. G\"om\"ory, {\it Supercond. Sci. Tech.} {\bf 10}, 523 (1997)
\bibitem{Gam90} P. L. Gammel, {\it J. Appl. Phys.} {\bf 67}, 4676 (1990)
\bibitem{Mal88} A. P. Malozemoff, T. K. Worthington, Y. Yeshurun, F. Holtzberg, P. H. Kes, {\it Phys. Rev. B} {\bf 38}, 7203 (1988)
\bibitem{Tin91} M. Tinkham, {\it Physica B} {\bf 169}, 66 (1991)
\bibitem{Bra91} E. H. Brandt, {\it Phys. Rev. Lett.} {\bf 67}, 2219 (1991)
\bibitem{Cof92} M. W. Coffey, J. R. Clem, {\it Phys. Rev. B} {\bf 45}, 9872 (1992)
\bibitem{Bra92} E. H. Brandt, {\it Phys. Rev. Lett.} {\bf 68}, 3769 (1992)
\bibitem{Kes89} P. H. Kes, J. Aarts, J. van der Berg, C. J. van der Beek, J. A. Mydosh, {\it Supercond. Sci. Tech.} {\bf 1}, 242 (1989)
\bibitem{Ges91} V. B. Geshkenbein, V. M. Vinokur, R. Fehrenbacher, {\it Phys. Rev. B} {\bf 43}, 3748 (1991)
\bibitem{Lin91} X. Ling, J. I. Budnick, in {\it Magnetic Susceptibility of Superconductors and Other Spin Systems} (edited by R. A. Hein, T. L. Francavilla, D. H. Liebenberg), Plenum Press, New York (1991), pages 377-388
\bibitem{Col41} K. S. Cole, R. H. Cole, {\it J. Chem. Phys.} {\bf 9}, 341 (1941)
\bibitem{Pow51} J. C. Powles, {\it Proc. Phys. Soc. B} {\bf 64}, 81 (1951)
\bibitem{Kub91} R. Kubo, M. Toda, N. Hashitsume, {\it Statistical Physics II: Nonequilibrium Statistical Mechanics}, Springer-Verlag Berlin (1991), pages 120-125
\bibitem{Emm91} J. H. P. M. Emmen, V. A. M. Brabers, W. J. M. de Jonge, {\it Physica C} {\bf 176}, 137 (1991)
\bibitem{Her97} T. Herzog, H. A. Radovan, P. Ziemann, E. H. Brandt, {\it Phys. Rev. B} {\bf 56}, 2871 (1997)
\bibitem{Hav66} S. Havriliak, S. Negami, {\it J. Polym. Sci. C} {\bf 14}, 99 (1966)
\bibitem{Cav84} R. J. Cava, R. M. Fleming, P. Littlewood, E. A. Rietman, L. F. Schneemeyer, R. G. Dunn, {\it Phys. Rev. B} {\bf 30}, 3228 (1984)
\bibitem{Mor01} A. H. Morrish, {\it The physical principles of magnetism}, IEEE Press, New York (2001), pages 87-101
\bibitem{Lee13} M. K. Lee, E. V. Charnaya, C. Tien, L. J. Chang, Y. A. Kumzerov, {\it J. Appl. Phys.} {\bf 113}, 113903 (2013)
\bibitem{Sta71} H. E. Stanley, {\it Introduction to Phase Transitions and Critical Phenomena}, Oxford University Press (1971)
\bibitem{Sag90} L. T. Sagdahl, S. Gjolmesli, T. Laegreid, K. Fossheim, W. Assmus, {\it Phys. Rev. B} {\bf 42}, 6797 (1990)
\bibitem{Cha11} M. Charilaou, J. F. L\"offler, A. U. Gehring, {\it Phys. Rev. B} {\bf 83}, 224414 (2011)
\bibitem{Hoh77} P. C. Hohenberg, B. I. Halperin, {\it Rev. Mod. Phys.} {\bf 49}, 435 (1977)
\bibitem{Pol08} M. Polichetti, M. G. Adesso, D. Zola, J. Luo, G. F. Chen, Z. Li, N. L. Wang, C. Noce, S. Pace, {\it Phys. Rev. B} {\bf 78}, 224523 (2008)
\bibitem{Ish90} T. Ishida, R. B. Goldfarb, {\it Phys. Rev. B} {\bf 41}, 8937 (1990)
\bibitem{Lam90} Q. H. Lam, Y. Kim, C. D. Jeffries, {\it Phys. Rev. B} {\bf 42}, 4846 (1990)
\bibitem{Pol00} M. Polichetti, M. G. Adesso, T. Di Matteo, A. Vecchione, S. Pace, {\it Physica C} {332}, 378 (2000)
\bibitem{Ade04} M. G. Adesso, C. Senatore, M. Polichetti, S. Pace, {\it Physica C} {404}, 289 (2004)
\bibitem{Tay07} B. J. Taylor, D. J. Scanderbeg, M. B. Maple, C. Kwon, Q. X. Jia, {\it Phys. Rev. B} {\bf 76}, 014518 (2007)
\bibitem{Gam91} P. L. Gammel, L. F. Schneemeyer, D. J. Bishop, {\it Phys. Rev. Lett.} {\bf 66}, 953 (1991)
\bibitem{Saf92} H. Safar, P. L. Gammel, D. J. Bishop, D. B. Mitzi, A. Kapitulnik, {\it Phys. Rev. Lett.} {\bf 68}, 2672 (1992)
\bibitem{Wil10} T. J. Williams, A. A. Aczel, E. Baggio-Saitovitch, S. L. Bud'ko, P. C. Canfield, J. P. Carlo, T. Goko, H. Kageyama, A. Kitada, J. Munevar, N. Ni, S. R. Saha, K. Kirschenbaum, J. Paglione, D. R. Sanchez-Candela, Y. J. Uemura, G. M. Luke, {\it Phys. Rev. B} {\bf 82}, 094512 (2010)
\bibitem{Tay07b} B. J. Taylor, M. B. Maple, {\it Phys. Rev. B} {\bf 76}, 184512 (2007)
\end{references}
\end{document}